\documentclass[aps,prb,manuscript,amsmath]{revtex4-2}
\usepackage{graphicx}
\usepackage{epsfig}
\usepackage{color}
\usepackage{bm}
\usepackage{float}

\UseRawInputEncoding

\begin{document}

\draft

\title{Engineering plasmon modes and their loss in armchair graphene nanoribbons by selected edge-extended defects}

\author{Thi-Nga Do$^{1}$, Po-Hsin Shih$^{1}\footnote{Corresponding author: {\em E-mail}: phshih@phys.ncku.edu.tw}$, Godfrey Gumbs$^{2}$, Danhong Huang$^{3}$}
\affiliation{$^{1}$ Department of Physics, National Cheng Kung University, Tainan 701, Taiwan \\
$^{2}$ Department of Physics and Astronomy, Hunter College of the City University of New York,
695 Park Avenue, New York, New York 10065, USA \\
$^{3}$ US Air Force Research Laboratory, Space Vehicles Directorate (AFRL/RVSU),
Kirtland Air Force Base, New Mexico 87117, USA
}

\date{\today}

\begin{abstract}

The effect of edge modification of armchair graphene nanoribbons (AGNRs) on the collective excitations are theoretically investigated. The tight-binding method is employed in conjunction with the dielectric function. Unconventional plasmon modes and their association with the flat bands of the specially designed AGNRs are thoroughly studied. We demonstrate the robust relationship between the novel collective excitations and both the type and period of the edge modification. Additionally, we reveal that the main features displayed in the (momentum, frequency)-phase diagrams for both single-particle and collective excitations of AGNRs can be efficiently tuned by edge-extended defects. Our obtained plasmon modes are found to be analogous to magnetoplasmons associated with collective excitations of Landau-quantized electrons. This work provides a unique way to engineer discrete magnetoplasmon-like modes of AGNRs in the absence of magnetic field.

\end{abstract}

\maketitle

\section{Introduction}
\label{sec1}

Graphene nanoribbons (GNRs) are quasi-one-dimensional systems that present great promise for nanoelectronics, optoelectronics, spintronics and other device applications. Nowadays, GNRs can be synthesized by using various experimental techniques, including either bottom-up \,\cite{bottomup1, bottomup2,bottomup3, bottomup4,bottomup5,bottomup6, bottomup7, bottomup8, bottomup9} or top-down\,\cite{topdown1, topdown2, topdown3, topdown4, topdown5, topdown6, topdown7} approaches. Edge defects, which usually appear during the fabrication processes for GNRs, have been shown to modify the fundamental characteristics of targeted materials. Therefore, special attention has been directed toward the edge-modified armchair graphene nanoribbons (AGNRs) because of their versatility, stability and interesting physics, e.g., topological behavior\,\cite{defect1, defect2}, band engineering\,\cite{bottomup2}, quantum-phase transition\,\cite{defect3}, and quantum magnetism\,\cite{defect4}.
\medskip

Collective excitation in a quantum many-body system becomes a critical concept for a deep understanding of the physical properties of materials, such as the  optical-absorption spectra\,\cite{optical}, fractional-quantum Hall plateaus\,\cite{fqhe}, electronic excitations\,\cite{dielectric} and decay rates\,\cite{decay,mdpi}, to name just a few. In this paper, we will concentrate on plasmons, which result from collective excitations of Coulomb-coupled charged carriers in the conduction or valence bands, and their various effects as well. Plasmon modes are found to be a quantum representative of the  charge-density oscillations in lattice structures, as shown in Fig.\,\ref{Fig1}$(a)$, for electrons at the valence-band extrema. In particular, effects of plasmons in graphene play important roles in their applications to optics\,\cite{app1}, nanolithography\,\cite{app2}, microscopy\,\cite{app3} and catalysis\,\cite{app4}. Electron energy loss spectroscopy is also known as one of the experimental techniques for probing collective electronic excitations in both long- and short-wavelength regimes\,\cite{eels1, eels2}. Meanwhile, plasmons have been shown to be greatly dependent on the lattice configuration\,\cite{lattice}, atomic interactions\,\cite{bilayerg}, doping\,\cite{doping}, temperature\,\cite{temp}, as well as external electric and magnetic fields\,\cite{sifield}. Specifically, magneto-plasmonics has already attracted a great deal of attention in the last decade because of its potential applications in relevant technologies\,\cite{tech1, tech2}. For quantum-well materials, magnetoplasmons are defined as the collective excitations between different quantized Landau levels (LLs), and they have been verified experimentally by infrared-optical absorption and inelastic-light scattering\,\cite{infrare1, infrare2}, as done for graphene\,\cite{expgraphene}. These unique magnetoplasmon modes have been theoretically investigated for layered graphene, doped graphene, GNRs, silicene and others\,\cite{sifield, layered, doped, GNR}.

\medskip

The influence of edge defects on electronic and transport properties of GNRs has been evaluated by various theoretical approaches, and it has also been demonstrated experimentally at the same time.  Cao et. al\,\cite{defect1} systematically investigated  different types of electronic topological phases in GNRs with verified widths, edges and end terminations. With respect to this effort, the effect of electronic correlations on the topological states in $7/9-$AGNR heterostructures was explored using a Green's function approach combined with an effective Hubbard Hamiltonian\,\cite{defect2} and the first-principles calculations\,\cite{bottomup2}. Moreover, the band structures and charge-density distributions of various types of edge-modified AGNRs were analyzed in Ref.\cite{defect3} by employing first-principles computations and the tight-binding model along with scanning-tunneling-spectroscopy measurements. Additionally, the quantum magnetism of topologically-designed GNRs was evaluated numerically based on
the Hubbard model\,\cite{defect4}. Furthermore, a study of the charge-transport mechanism of GNRs was conducted in Ref.\cite{defect5} based on a semi-classical model with the tight-binding approximation. Despite recent progress in this field,   there is still a  lack of a clear understanding about the correspondence between edge-modification and electron-electron interaction in GNRs. Therefore, a comprehensive investigation is now needed for physics explanations to experimentally-observed electronic and plasmon-excitation properties in edge-defect AGNRs.
\medskip

In the present work, we assess the influence of edge-extended defects on the collective excitations in AGNRs. The nontrivial plasmon mode dispersion relation, associated with nearly-flat bands of specifically designed AGNRs, are carefully explored. Our research reveals that the plasmon spectra in AGNRs  strongly depend on the type and period of the edge modifications. Very importantly, this work presents a prescription for creating quantized plasmon modes in the absence of magnetic field. Numerically, we employ a generalized tight-binding model to calculate the energy bands and (momentum, frequency)-phase diagrams for plasmon excitations. The optical-selection rules for inter-band transitions between two flat bands and related damping are analyzed fully in order to clarify its analogy to magnetoplasmons from quantized LLs.  Meanwhile, we also discuss the relationship between our calculated results with edge and one-dimensional (1D) plasmas.
\medskip

The rest of the paper is organized as follows. As a beginning, in Sec.\ \ref{sec2} we first present our method for conducting the numerical computations, including edge-extended defects in graphene nanoribbons. In Sec.\ \ref{sec3}, we present a detailed discussion for plasmon excitation in graphene nanoribbons, and meanwhile, introduce an effective dielectric function for this system. Numerical results and discussion are presented in Sec.\ \ref{sec4}.  Finally, conclusions drawn from this paper are summarized in Sec.\ \ref{sec5}.

\section{Tight-binding model}
\label{sec2}

\begin{figure}[ht]
\begin{center}
\includegraphics[width=1.0\linewidth]{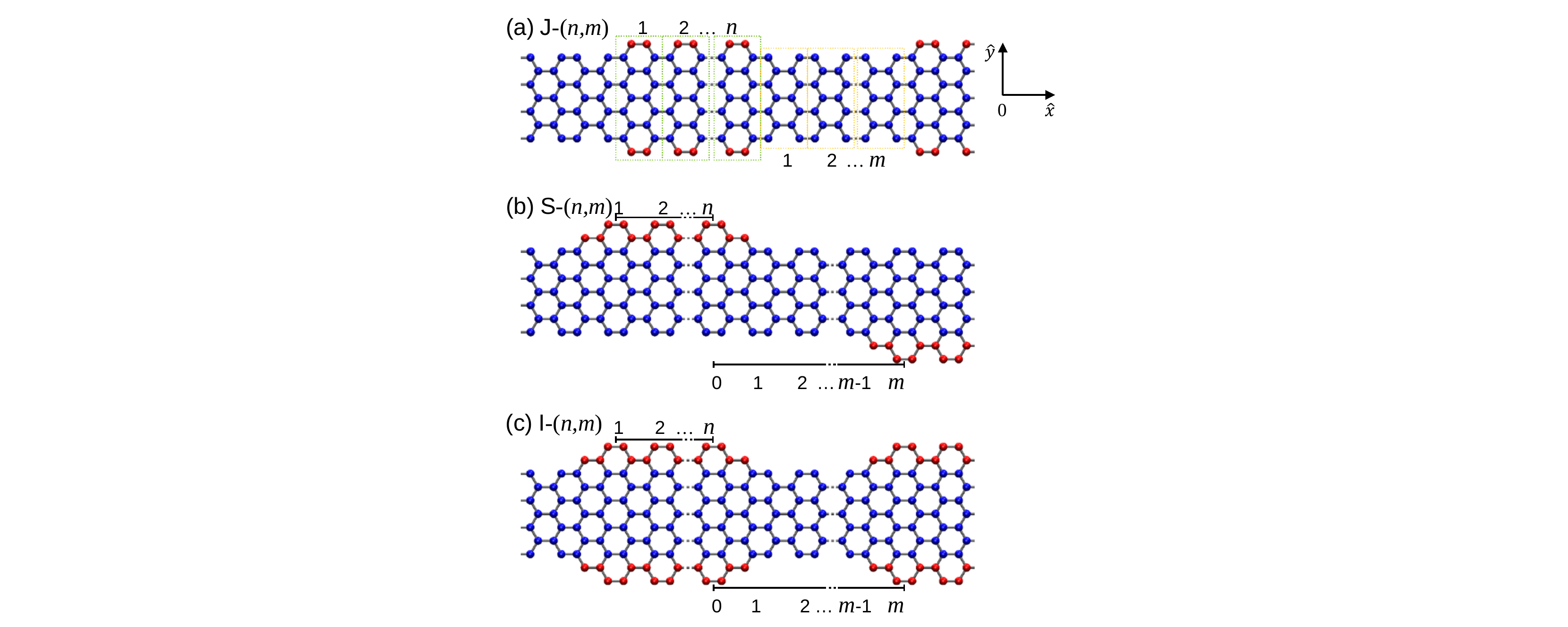}
\end{center}
\caption{(color online) Lattice structures of three different types of edge-modified AGNRs: $(a)$ $7/9$-AGNR heterojunction; $(b)$ staggered edge-extended AGNR, and $(c)$ inline edge-extended AGNR.  The blue and red spheres represent carbon atoms within the 7-AGNR and those along the extended edges.}
\label{Fig1}
\end{figure}

For pristine GNR, the tight-binding model mainly considers the hoping interaction between nearest-neighbor atoms. The width of an AGNR, denoted as N, is defined as the number of atoms across  the transverse direction of the ribbon. The edge-extended defect is introduced into an N-AGNR by creating a periodic sequence of extended AGNR segments on both sides along the ribbon backbone.  Consequently, the tight-binding Hamiltonian needs to be modified accordingly in order to take into account the effect due to an extended edge. Here, we consider three types of edge-modified AGNRs, namely $7/9-$AGNR heterojunction, staggered edge-extended AGNR, and inline edge-extended AGNR, as shown in Figs.\,\ref{Fig1}$(a)$ through $\ref{Fig1}(c)$. These three edge configurations are denoted as J-$(n,m)$, S-$(n,m)$ and I-$(n,m)$, respectively. For J-$(n,m)$, $n$ and $m$ correspond to the number of primitive unit cells for the $9-$AGNR and $7-$AGNR. On the other hand, $n$ stands for the length of the edge-extended segment whereas $m$  represents the separation between two segments in S-$(n,m)$ and I-$(n,m)$ systems. Here, the indices $n$ and $m$ are associated with the real-space distance in units of $3b$ with $b=1.42\,$\AA\ being the C-C bond length. The hetero-junction J-$(n,m)$ is formed by the combination of a $N$-AGNR and a $(N+2)$-AGNR in a way that ensures the equivalence between two ribbon edges. In a S-$(n,m)$ AGNR, however, the extended-edge segments are $(N+2)$ wide but the extension is alternate  on each side of the ribbon. Finally, for a I-$(n,m)$ AGNR, the $(N+4)$ ribbon segments are located symmetrically along the backbone.
\medskip

The $p_z$-orbital tight-binding Hamiltonian of an edge-modified AGNR can be written as

\begin{equation}
{\cal H} =  \sum\limits_{\langle i,j \rangle}\,\gamma^{R_{ij}} C_{i}^{\dag} C_{j} + h.c.\ ,
\label{eqn-1}
\end{equation}
where $C_{i}^{\dag}$ and $C_{j}$ are operators which create and annihilate an electronic state at the lattice sites $i$ and $j$, respectively. It should be noticed that, $i$ and $j$ are properly selected so that the Hamiltonian satisfies the boundary condition related to the ribbon width. $\gamma^{R_{ij}}=2.6\,$eV is the nearest-neighbor hopping integral with $R_{ij}$ being the translation vector between two atoms\,\cite{GNR}, and $h.c.$ represents the Hermitian conjugate of the first term. It should be noted that, the Hamiltonian in Eq.\,\eqref{eqn-1} includes the boundary condition of the AGNR through the hopping interaction and relevant phase terms. Therefore, the corresponding band structure and wave functions automatically satisfy such the boundary condition.
\medskip

\section{Dielectric function}
\label{sec3}

When an AGNR is subjected to an incident electron beam, both the valence and conduction electrons will experience a charge redistribution, leading to a polarization field in response to an external  potential and producing a dynamical screening to this external field at the same time. Consequently, the effective potential between two charges becomes sum of the external and induced potentials. For an ideal graphene nanoribbon without edge defects, by using the standard many-body theory,\,\cite{book} the dielectric-function tensor within the random-phase approximation can be generally written as

\begin{equation}
\epsilon_{jm,\,j'm'}^{\mu\mu',\,hh'}(q_x,\omega)=\delta_{h,\mu}\delta_{h',\mu'}\delta_{j,j'}\delta_{m,m'}
-V_{jm,\,j'm'}^{\mu\mu',\,hh'}(q_x)\,\chi^{hh'}_{j'm'}(q_x,\omega)\ ,
\label{eqn2}
\end{equation}
where $q_x$ is the longitudinal transition wave number along a nanoribbon, $\omega$ is the angular frequency of a testing field, the index set $\{h,h';\,\mu,\mu'\}=c,v$ labels the conduction or valence band around $K$ and $K'$ valleys, while the other index set $\{j,m;\,j',m'\}=1,\,2,\,\cdots$ labels split subbands within either a conduction or a valence band due to lateral quantization of the nanoribbon. The plasmon modes of the system can be computed from the determinant of  dielectric-function tensor in Eq.\,\eqref{eqn2}, i.e., ${\cal D}et\left[\epsilon_{jm,\,j'm'}^{\mu\mu',\,hh'}(q_x,\omega)\right]=0$. Here, the diagonal matrix elements in Eq.\,\eqref{eqn2} give rise to the dispersion of individual plasmon modes, while the off-diagonal matrix elements in Eq.\,\eqref{eqn2} describe the couplings between different plasmon modes.
\medskip

In Eq.\,\eqref{eqn2}, we have introduced the subband polarization function $\chi^{hh'}_{j'm'}(q_x,\omega)$, defined as

\begin{equation}
\chi^{hh'}_{j'm'}(q_x,\omega)=\frac{g_s}{2\pi}\,\int\limits_{1^{st}\,{\rm BZ}} dk_x\,\left[\frac{f_0(\varepsilon_{k_x,m'}^{h})-f_0(\varepsilon_{k_x+q_x,j'}^{h'})}
{\hbar(\omega+i\delta)-\varepsilon_{k_x+q_x,j'}^{h'}+\varepsilon_{k_x,m'}^{h}}\right]\,{\cal F}_{j'm'}^{hh'}(k_x,\,q_x)\ ,
\label{eqn3}
\end{equation}
where the integral with respect to wave number $k_x$ is limited to the first Brillouin zone, $g_s=2$ takes into account the spin degeneracy, $\delta$ = 0.01 eV is the energy width due to various deexcitation mechanisms, $\delta\ll\omega$ is associated with a homogeneous diagonal-dephasing rate of electrons, $\varepsilon_{k_x,m'}^{h}$ is the subband energy, $f_0(x)=\{1+\exp[(x-u_0)/k_BT)]\}^{-1}$ is the Fermi function for thermal-equilibrium electrons, $T$ is the temperature, and $u_0(T)$ is the chemical potential of electrons.
For graphene nanoribbons even with symmetrical edge-extended defects, the mirror symmetry between conduction and valence bands can be maintained, and then we have $u_0(T)=0$. Moreover, the dimensionless form factor ${\cal F}_{j'm'}^{hh'}(k_x,\,q_x)$ in Eq.\,\eqref{eqn2} for longitudunal wave functions is calculated as\,\cite{mdpi,GNR,dielectric}

\begin{equation}
{\cal F}_{j'm'}^{hh'}(k_x,\,q_x)=\left|\langle k_x + q_x;j',h'\left|\texttt{e}^{iq_xx}\right| k_x;m',h\rangle\right|^{2}\ .
\label{eqn4}
\end{equation}

\medskip

On the other hand, the Coulomb-interaction matrix in Eq.\,\eqref{eqn2} is formally written as

\begin{eqnarray}
\nonumber
V_{jm,\,j'm'}^{\mu\mu',\,hh'}(q_x)&=&\frac{e^2}{2\epsilon_0\epsilon_b}\,\int\limits_{-\infty}^{\infty}\,\frac{dp}{\sqrt{q_x^2+p^2}}\,\\
&\times&\int\limits_0^W dy\,\int\limits_0^W dy'
[\,\psi^{\mu'}_j(y)]^*[\,\psi^{h'}_{j'}(y')]^*\texttt{e}^{ip(y-y')}\,\psi^{h}_{m'}(y')\,\psi^\mu_m(y)\ ,
\label{eqn5}
\end{eqnarray}
which is dependent on subband indexes $\{j,j',m,m'\}$ and on band indexes $\{h,h',\mu,\mu'\}$.
In Eq.\,\eqref{eqn5}, $\epsilon_0$ is the vacuum permittivity, $\epsilon_b$ is the dielectric constant of nanoribbon material, $W$ stands for the width of nanoribbon, and $\psi^h_j(y)$ is the wave function.
\medskip

Physically, one should note that the dimensionless form factor ${\cal F}_{j'm'}^{hh'}(k_x,\,q_x)$ in Eq.\,\eqref{eqn4} is only contained in the subband polarization function $\chi^{hh'}_{j'm'}(q_x,\omega)$ but not in Eq.\,\eqref{eqn3} for the Coulomb-interaction matrix. In addition, the broken transnational symmetry in the transverse direction of a nanoribbon leads to unique dependence of the form factor on the subband indices $j'$ and $m'$.
By neglecting weak off-diagonal matrix elements for the coupling between different plasmon modes, we can simply set $h=\mu$, $h'=\mu'$, $j=j'$ and $m=m'$ for the second term in Eq.\,\eqref{eqn2}, and therefore, we get the so-called effective dynamical dielectric function $\epsilon_{\rm eff}(q_x,\omega)$ defined as

\begin{eqnarray}
\begin{aligned}
\epsilon_{\rm eff}(q_x,\omega)&\equiv\epsilon_1+i\epsilon_2=
1-\sum\limits_{h,h'=c,v}\,\sum\limits_{j',m'}\,
\int\limits_{1st\,BZ}\frac{2dk_x}{2\pi}\,{\cal F}_{j'm'}^{hh'}(k_x,\,q_x)\,U^{hh'}_{j'm'}(k_x,\,q_x)\\
&\times \left[\frac{f(\varepsilon_{j'}^{h'}(k_x + q_x))-f(\varepsilon_{m'}^{h}(k_x))}
{\varepsilon_{j'}^{h'}(k_x + q_x)-\varepsilon_{m'}^{h}(k_x)-\hbar(\omega+i\delta)}\right]\ ,
\end{aligned}
\label{eqn6}
\end{eqnarray}
where the Coulomb-interaction matrix in Eq.\,\eqref{eqn6} is calculated as

\begin{equation}
U^{hh'}_{j'm'}(k_x,\,q_x)=\frac{e^2}{4\epsilon_0\epsilon_b}\int\limits_{-\infty}^\infty \frac{dp}{\sqrt{q_x^2+p^2}}\,\left|{\cal Q}^{hh'}_{j'm'}(k_x,q_x;\,p)+\left[{\cal Q}^{hh'}_{j'm'}(k_x,q_x;\,-p)\right]^*\right|^2\ ,
\label{eqn7}
\end{equation}
while the form factor for transverse wave functions takes the form

\begin{equation}
{\cal Q}^{hh'}_{j'm'}(k_x,q_x;\,p)=\langle k_x+q_x;j',h'\left|\texttt{e}^{ipy}\right|k_x;m',h\rangle\ .
\label{eqn8}
\end{equation}

\section{Results and Discussion}
 \label{sec4}

Edge modification is an efficient method for manipulating electronic band-structures of AGNRs. By introducing edge extension into an AGNR, the energy bands in the vicinity of the zero energy are governed by the edge states \cite{defect3}. It was shown for the pristine AGNRs that, the band gap is inversely proportional to the ribbon width \cite{AGNR}. Here, we consider the narrow ribbons with relatively large band gap so that the effect of the edge-defect on the electronic properties of the system is manifest. In particular, the existence of the edge bands at low energy is visible without the interfering of other irrelevant bands. This allows the further study of the (momentum, frequency)-phase diagrams, plasmon spectra and their dependence on the edge-defect.
Moreover, our theoretical prediction can be compared with the experimental measurements which have only been focused on the small-width GNRs up to this point of time \cite{bottomup2, defect3}.
It is worth mentioning that, the nanoribbon truncated TBM has a built-in ribbon width and does not require a specific ribbon width for its subband quantization, and the evaluation of nanoribbon width in this work is just for indication and comparison purposes. \\

\begin{figure}[ht]
\begin{center}
\includegraphics[width=1.0\linewidth]{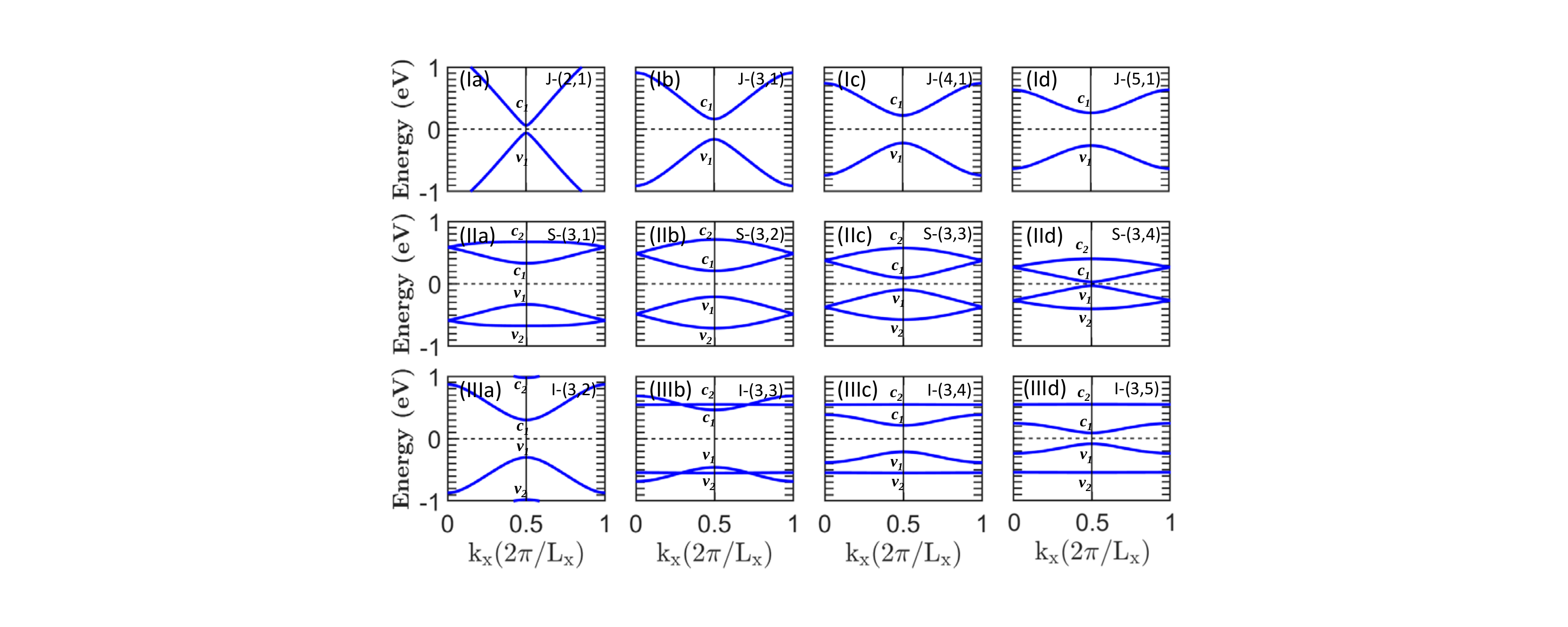}
\end{center}
\caption{(color online)  Calculated band structures for edge-modified AGNRs along the $x$ direction as functions of wave number $k_x$ in the whole first Brillouin zone, where
$c_{1,2}$ and $v_{1,2}$ label the conduction and valence bands, respectively.  The labels I, II and III refer to J-$(n,m)$, S-$(n,m)$ and I-$(n,m)$ edge-extended defects, respectively.}
\label{Fig2}
\end{figure}

The band gap and energy dispersion of AGNRs depend significantly on the selected edge configuration labeled denoted by $(n,m)$. The low-lying valence and conduction edge bands change substantially by varying $(n,m)$, as displayed in Fig.\,\ref{Fig2} for three different types of edge-modification. First, for $7/9$-hetero-junctions, there exists a pair of symmetric parabolic bands $(v_1,c_1)$ in the vicinity of the Fermi level $E_F = 0$. The band gap between the $v_1$ and $c_1$ bands can be adjusted  gradually by changing either $n$ or $m$. As an example, by increasing $n$ from $1$ to $4$ for the J-$(n,3)$ systems, the band gap is widened, as seen in Figs.\,\ref{Fig2}(I$a$)-\ref{Fig2}(I$d$).   Additionally, the energy bands for J-$(n,3)$ structures become flattened when $n$ is increased, which could lead to novel features for electronic excitations in these systems.  Next, when AGNRs are subjected to S-$(3,m)$ defects, instead, there appear two pairs of valence and conduction bands near $E_F = 0$, i.e. $(v_1,c_1)$ and $(v_2,c_2)$. In this case,  both the conduction and valence bands are found to be doubly degenerate always at the first Brillouin-zone boundaries ($k_x=0$ and $2\pi/L_x$) regardless of the S-$(3,m)$ modifications. Specifically, the band gap for a S-$(3,m)$ defect, defined as the separation between the $v_1$ and $c_1$ bands, is enhanced as $m$ is increased from $1$ to $4$, as displayed in Figs.\,\ref{Fig2}(II$a$) through \ref{Fig2}(II$d$). For such a modification scheme with $m$, the group velocities (or slopes) of these four bands only change slightly. Finally, the alteration of the energy dispersion for the I-$(n,m)$ defects becomes even more visible. Figures\ \ref{Fig2}(III$a$)-\ref{Fig2}(III$e$) illustrate the features of two low-energy bands corresponding to fluctuations with $n=3$ or $4$ and $m$ running from $2$ to $5$. For I-$(n,m)$ defects, the band gap between the $v_1$ and $c_1$ bands is continuously tuned by choosing different edge configurations. The inter-subband spacing between two valence ($v_1,\,v_2$) or two conduction ($c_1,\,c_2$) bands also increases or decreases with $n$ and/or $m$ values non-monotonically, as demonstrated by Figs.\,\ref{Fig2}(III$c$) for I-$(3,4)$, \ref{Fig2}(III$d$) for I-$(3,5)$ and \ref{Fig2}(III$e$) for I-$(4,4)$.   The ($v_1,\,v_2$) or ($c_1,\,c_2$) bands for I-$(3,3)$ can even cross each other or otherwise get separated from each other within the whole first Brillouin zone.  Interestingly, the pair of ($v_2,\,c_2$) bands are found to be completely flat, which resembles magnetically quantized LLs. Such flat bands can survive despite increasing $n$ or $m$, as can be verified from Figs.\,\ref{Fig2}(III$c$)-\ref{Fig2}(III$d$). Here, it is worth emphasizing that there exist band-gap closing and opening for specifically selected choices for ($n,m$), such as S-$(2,3)$, I-$(2,5)$ and J-$(2,4)$ AGNRs (see Supplemental Materials for details).
\medskip

\begin{figure}[ht]
\centering
{\includegraphics[width=1.0\linewidth]{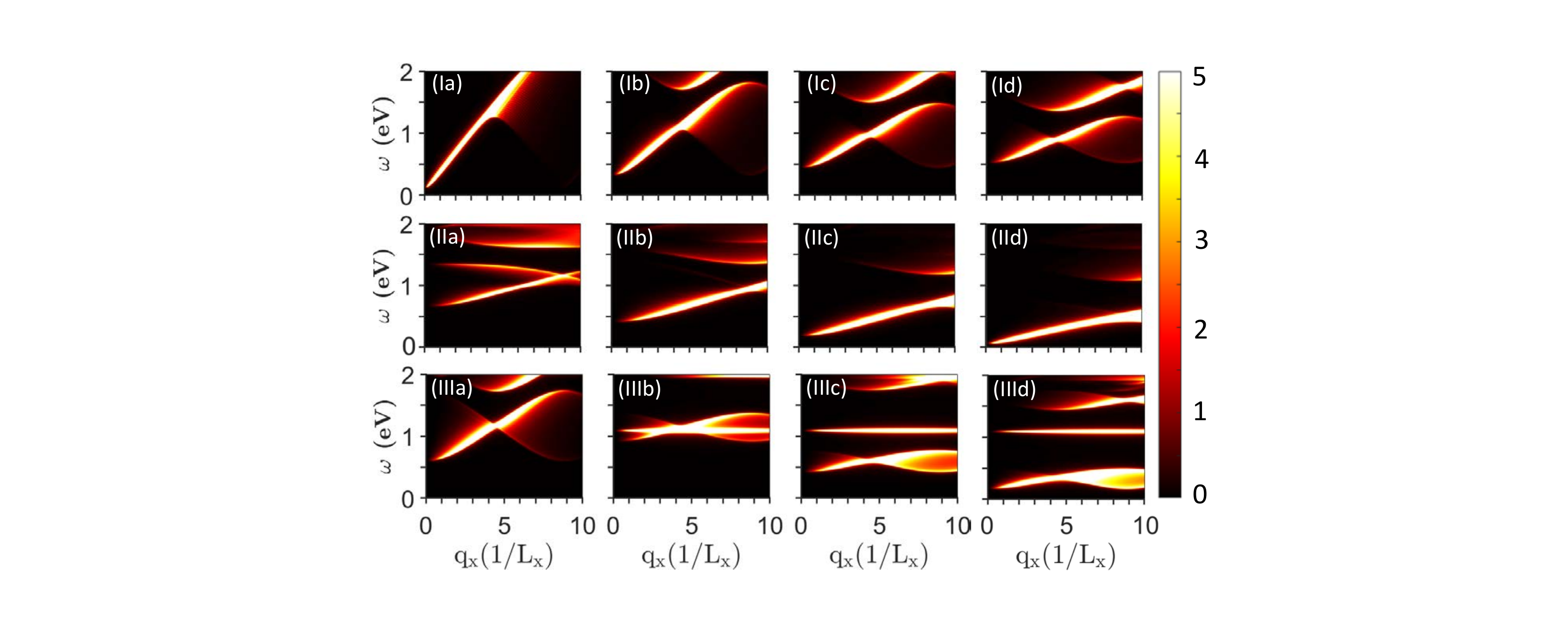}}
\caption{(color online) The $(q_x,\omega)$-phase diagram of SPEs for AGNRs with three different edge modifications.  The assignment of panels to three different edge-defect configurations is kept the same as those in Fig.\,\ref{Fig2} for easy comparison.}
\label{Fig3}
\end{figure}

Engineering the band structure can lead to fantastic properties of electrons in the system as well as exotic excitation phenomena. The AGNRs exhibit both single-particle excitations (SPEs) and collective excitations.  Generally, the SPEs, which are described by the imaginary part ($\epsilon_2$) of the dielectric function in Eq.\,\eqref{eqn6}, include contributions from both intraband and interband transitions.  However,  in the present study, we only consider $E_F = 0$ for pristine graphene, and therefore, intraband excitations are forbidden in this case.  Physically, SPEs can be characterized as a loss channel for plasmon excitation in the system, which is related to the real part of the  optical conductivity determined by the  Kubo formula.\,\cite{fqhe}
The SPE loss is known to depend on the transferred momentum and the Fermi energy for doped graphene. Our calculations further reveal that SPEs are also affected by lattice configurations.
Figure\ \ref{Fig3} displays the $(q_x,\omega)$-phase diagram of SPEs for various edge-modified AGNRs. From Fig.\,\ref{Fig2}, we know that the simplest case among all panels in Fig.\,\ref{Fig3} is Fig.\,\ref{Fig3}(I$a$) for J-$(1,3)$, corresponding to SPEs between a pair of occupied and unoccupied parabolic bands. The bright region for strong SPEs extend from low to high momentum $\hbar q_x$ and frequency $\omega$, exhibiting a linear relationship between them.  Furthermore, the starting point at $q_x=0$ corresponds to the band gap and gives rise to a value $\hbar\omega=0.2\,$eV. The bright region becomes wider and ended with an extended faint area for sufficiently high momentum ($q_x \geq$ 5/$L_x$) and frequency ($\hbar\omega \leq 1\,$eV) where its overlap with weakened interband excitation occurs. With increasing $n$ for J-$(n,3)$, we find two bright regions for substantial SPEs in the range of $0\leq\hbar\omega \leq 2\,$eV, as seen in Figs.\,\ref{Fig3}(I$b$)-\ref{Fig3}(I$d$). Here, the upper high intensity region results from the electronic transitions from $v_1$ to other higher conduction bands (not shown) and from the lower valence subbands (not shown) to the $c_1$ band. Clearly, the regions for strong SPEs show up within a narrower range of $\omega$ for large values of $n$, as can be seen from the reduced group velocity in different columns of the first row in Fig.\,\ref{Fig2}.
\medskip

The SPEs of S-$(3,m)$ AGNRs comprise a rather complex  $(q_x,\omega)$-phase diagram, as demonstrated in  Figs.\,\ref{Fig3}(II$a$) through \ref{Fig3}(II$d$). The electronic excitations can be classified via three transition channels corresponding to $v_1 \to c_1$, ($v_1 \to c_2$, $v_2 \to c_1$), and $v_2 \to c_2$. These transition channels exhibit distinctive distributions of substantial SPEs which diversify the $(q_x,\omega)$-phase diagram. For example, the S-(3,1) system presents three clear traces of strong SPEs, as shown in Fig.\,\ref{Fig3}(II$a$). From bottom to top, the first, second, and third bright regions are related to the transitions of $v_1 \to c_1$, ($v_1 \to c_2$, $v_2 \to c_1$), and $v_2 \to c_2$, respectively.
With increasing $q_x$, the bottom SPE goes up with frequency, while the opposite is true for the middle SPE. Especially, the top SPE appears flat, which reflects the flat bands of $v_2$ and $c_2$. When the configuration is varied by increasing $m$, the phase diagram is drastically modified, as shown in Figs.\,\ref{Fig3}(II$b$) through Fig.\,\ref{Fig3}(II$d$), in which only the bottom and top SPEs survive since the transition for the middle SPE becomes forbidden by broken mirror symmetry of the energy bands in Figs.\,\ref{Fig2}(II$b$)  through Fig.\,\ref{Fig2}(II$d$). This selection rule for SPEs also applies to other  S-$(3,m)$ defect systems. Here, both the bottom and top SPEs are shifted down in frequency, and the bottom SPE is enhanced remarkably and extends to lower $q_x$. On the contrary,  the intensity of the  top SPE is weakened and eventually faded away in the vicinity of $q_x= 0$.
\medskip

More exotic features of the $(q_x,\omega)$-phase diagram for SPEs can be found from the I-$(n,m)$ AGNRs. Figure\ \ref{Fig3}(III$a$)  through \ref{Fig3}(III$e$) demonstrate the fluctuation of SPEs with the variation of $n$ and $m$. In this case, the main features of SPEs become very sensitive to the change of defect configurations $(n,m)$. For the I-$(3,2)$ system, the phase diagram presents two non-overlapped SPEs, similar to that for J-$(4,3)$ in Fig.\,\ref{Fig3}(I$d$), due to almost the same energy dispersion. However, the I-$(3,3)$ acquires three strong SPEs, in which two lower SPEs are partially overlapped with each other to display a unique shape in Fig.\,\ref{Fig3}(III$b$). This overlap can be attributed to the crossing valance and conduction bands in Fig.\,\ref{Fig2}(III$b$). Surprisingly, there appears a flat SPE around $\omega \sim 1\,$eV across a wide range of $q_x$, corresponding to the transition between two flat bands $v_2$ and $c_2$ in Fig.\,\ref{Fig2}(III$c$). Such a flat SPE is also seen for the I-$(3,4)$, I-$(3,5)$, and I-$(4,4)$ AGNRs, referring to Figs.\,\ref{Fig3}(III$c$)  through \ref{Fig3}(III$d$), for which three substantial SPEs are found to be well separated and the spacing between them varies with different defects $(n,m)$. In short, all the features in the $(q_x,\omega)$-phase diagrams for various SPEs can be completely described by the corresponding energy dispersion of the systems. We have elucidated a critical bridge connecting  plasmon and single-particle excitations with band structures in AGNRs.
\medskip

\begin{figure}[ht]
\centering
{\includegraphics[width=1.0\linewidth]{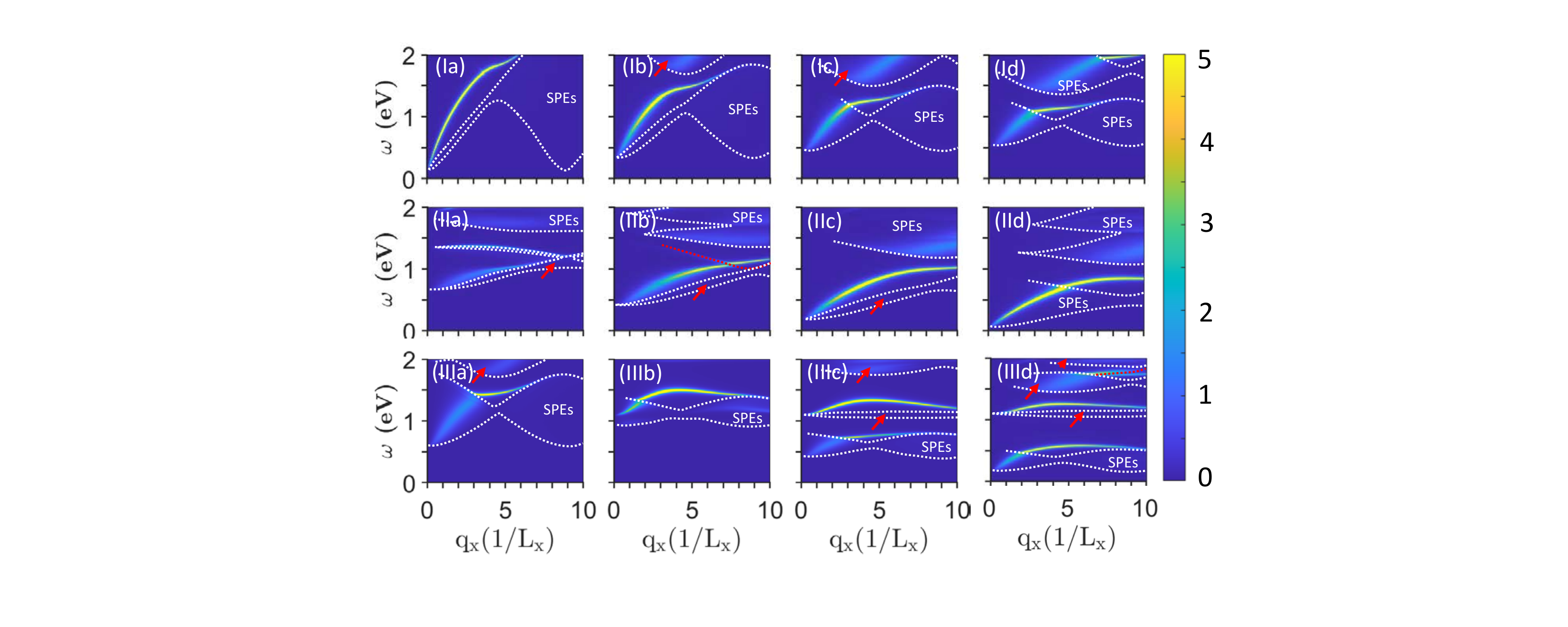}}
\caption{(color online) The $(q_x,\omega)$-phase diagram for plasmon spectra with three different edge modifications. Here, the SPE boundaries are marked by white dashed curves.
In addition, the red dashed curves and arrows indicate the SPEs regions. Here, the assignment of panels to three different edge-defect configurations is  kept the same as those in Fig.\,\ref{Fig2} for easy comparisons.}
\label{Fig4}
\end{figure}

After a discussion about features of SPEs in AGNRs with different defect configurations, we now turn our attention to plasmon excitation (PLE). Physically, PLEs represent self-sustaining induced charge-density oscillations and result from a depolarization process from photo-excited charged carriers within a system for responding to a testing optical field, which appears as a dielectric function for quantifying the screening to interactions with these charged carriers.   We investigate the dynamic charge screening related to calculated energy bands within a wide frequency range in order to demonstrate full features in the $(q_x,\omega)$-phase diagrams.   Figure\ \ref{Fig4} presents PLEs for various edge-extended AGNRs in terms of their momentum and frequency dependence. The boundaries of SPEs are also presented  in order to address the loss channel PLEs.  As seen in Fig.\,\ref{Fig4}, the PLE modes can be efficiently controlled by modifying edge defects, and they exhibit distinct features in terms of frequency range, momentum distribution, peak intensity, and number of PLE modes and their continuities.  For the J-$(n,3)$ AGNRs, the PLE appears as a single mode going up in frequency with increasing $q_x$ [see J-$(1,3)$, J-$(2,3)$, J-$(3,3)$] or multiple modes [see J-$(4,3)$].
With increasing $n$, the PLE becomes non-dissipating only within a large $q_x$ and small $\omega$ region. Especially, the PLE mode appears dispersionless for J-$(4,3)$. Meanwhile, the PLE modes can be either continuous without being influenced by SPEs [see Figs.\,\ref{Fig4}(I$a$) and \ref{Fig4}(I$b$)] or broken [see Fig.\,\ref{Fig4}(I$c$)] or even damped out by SPEs [see Fig.\,\ref{Fig4}(I$d$)].   We further find that the PLEs occur in broader $q_x$ range for S-$(3,m)$ and I-$(n,m)$ AGNRs compared to J-$(n,3)$ AGNR. Among three different types of edge-extended defects, the cutoff of PLE mode by SPE is more visible for the cases of S-$(1,3)$ and S-$(4,3)$ [see Figs.\,\ref{Fig4}(II$a$) and 4(II$d$)]. Moreover, the enhanced Landau damping to PLEs by SPEs can be seen for I-$(3,2)$, I-$(3,3)$ and I-$(3,5)$ systems, as shown in Figs.\,\ref{Fig4}(III$a$), \ref{Fig4}(III$b$) and \ref{Fig4}(III$d$), respectively.
\medskip

Specifically, the $(q_x,\omega)$-dependence of PLEs is peculiar for I-$(n,m)$ systems, as presented in Figs.\ref{Fig4}(III$a$) through \ref{Fig4}(III$d$). Here, the non-dissipating PLE regions are limited to a narrow range of $\omega$ around the boundaries of infrared and visible SPE regions throughout the $0\leq q_x\leq 10/L_x$ range. The PLEs exhibit a unique parabolic shape, implying a complicated relationship between their frequency and transferred momentum. Interestingly, the multi-PLE modes come to exist for the edge-defect systems with appropriate values of $(n,m)$, e.g., I-$(3,5)$ [see Fig.\,\ref{Fig4}(III$d$)]. Totally, there are three PLE branches in the range of $0\leq\hbar\omega\leq 2\,$eV. However, the bottom and top ones of the I-$(3,5)$ system are partially suppressed by the SPEs, while the strong middle one become dissipation free.   Additionally, these observed discrete multi-PLE modes resemble magnetoplasmons under a perpendicular magnetic field\,\cite{sifield}.   Therefore, it is clear that the PLEs, corresponding to the flat bands, behave very much like PLEs of LLs.   Our theoretical results in this paper pave the way to search for suitable materials with technological applications but without the need for a magnetic field.
\medskip

The edge plasmon is an important fundamental characteristic of GNRs which has been verified by the nano-infrared imaging \cite{edgep1, edgep2}.
Edge plasmons have been demonstrated to be the lowest loss plasmon mode in GNRs\,\cite{edgep3}, thus considered as the best candidate for the graphene plasmon based device applications. They arise from the energy localized on the edge of the ribbon.
It was predicted that the edge plasmons in GNRs appeared as either symmetric or asymmetric modes are propagating along the ribbon backbone, furthermore, they are strongly influenced by the edge defects\,\cite{edgep4}.
Our observed quantized plasmon modes and their dependence on the edge-modification are in a good agreement with the previous experimental and theoretical results.

\section{Concluding Remarks}
\label{sec5}

In conclusion, using the tight-binding model, we have investigated the exotic plasmons within edge-extended AGNRs. The $(q_x,\omega)$-phase diagrams for both SPEs and PLEs are calculated numerically from the explicit form of dielectric function within the random-phase approximation. The robust dependence of band structure on the configuration of edge-defect systems has been discussed thoroughly.  Moreover, we have revealed the full features of SPEs as well as their connection to the calculated band structures, in addition to their roles in dissipation of PLEs. We have further demonstrated extraordinary features of PLE modes in momentum-frequency space for three different types of edge-modified AGNRs, including the multi-PLE modes which resemble the magnetoplasmons resulting from quantized LLs. Finally, our theoretical results suggest an efficient way to engineer PLEs in AGNRs by introducing selected edge-defect configurations. Meanwhile, this work also enables the generation of discrete PLE modes without the use of an external magnetic field, which could be useful for applications to the development of next-generation  optical and transport quantum devices.

\begin{acknowledgements}
P. -H. Shih would like to thank the Ministry of Science and Technology of Taiwan for the support through the Grant No. MOST 110-2636-M-006-002.   G.G. would like to acknowledge the support from the Air Force Research Laboratory (AFRL)
through Grant No. FA9453-21-1-0046  D.H. would like to acknowledge the financial support from the Air Force Office of Scientific Research (AFOSR). T. -N. Do would like to thank the NCKU 90 and Beyond project for the support.
\end{acknowledgements}

\end{document}